\documentclass[lettersize,journal]{IEEEtran}
\usepackage{amsmath,amsfonts}

\usepackage[linesnumbered, ruled]{algorithm2e}
\SetKwRepeat{Do}{do}{while}%
\usepackage{cite}
\usepackage{array}
\usepackage{soul}

\usepackage{tabularx}
\usepackage{booktabs}

\usepackage[caption=false,font=normalsize,labelfont=sf,textfont=sf]{subfig}
\usepackage{algorithmic}
\usepackage{textcomp}
\usepackage{stfloats}
\usepackage{url}
\usepackage{verbatim}
\usepackage{graphicx}
\usepackage[T1]{fontenc} %摘要字体加粗
\usepackage{amssymb} % 三角等号
\usepackage[backref=False]{hyperref}
\def\BibTeX{{\rm B\kern-.05em{\sc i\kern-.025em b}\kern-.08em
    T\kern-.1667em\lower.7ex\hbox{E}\kern-.125emX}}
\usepackage{balance}
\newcommand{\Rmnum}[1]{\uppercase\expandafter{\romannumeral #1}}  %定义命令输入大写罗马数字
\usepackage{tikz,xcolor,hyperref}

\definecolor{lime}{HTML}{A6CE39}
\DeclareRobustCommand{\orcidicon}{
	\begin{tikzpicture}
		\draw[lime, fill=lime] (0,0)
		circle[radius=0.16]
		node[white]{{\fontfamily{qag}\selectfont \tiny \.{I}D}}; 
	\end{tikzpicture}
	\hspace{-2mm}
}
\foreach \x in {A, ..., Z}{%
	\expandafter\xdef\csname orcid\x\endcsname{\noexpand\href{https://orcid.org/\csname orcidauthor\x\endcsname}{\noexpand\orcidicon}}
}

\begin{document}
\title{Towards Governance of Localized VANET: An Adjustable Degree Distribution Model}
\author{Ruixing~Ren\hspace{-1.5mm}\orcidA{}, Junhui~Zhao\hspace{-1.5mm}\orcidB{},~\IEEEmembership{Senior~Member,~IEEE}, Xiaoke~Sun\hspace{-1.5mm}\orcidC{}, and Shanjin~Ni\hspace{-1.5mm}\orcidD{}
\thanks{Corresponding author: Junhui Zhao.

Ruixing Ren, Junhui Zhao are with the School of Electronic and Information Engineering, Beijing Jiaotong University, Beijing 100044, China. (e-mail: renruixing0604@163.com; junhuizhao@hotmail.com)

Xiaoke Sun and Shanjin Ni are with the National Computer Network Emergency Response Technical Team/Coordination Center of China (CNCERT/CC), Beijing 100029, China.}
}

\maketitle

\begin{abstract}
Vehicular Ad-hoc Networks (VANETs) serve as a critical enabler for intelligent transportation systems. However, their practical deployment faces a core governance dilemma: the network topology requires a dynamic trade-off between robustness against targeted attacks and ensuring low-latency information transmission. Most existing models generate fixed degree distributions, lacking the ability to adapt autonomously to the demands of diverse traffic scenarios. To address this challenge, this paper innovatively proposes a schedulable degree distribution model for localized VANETs. The core of this model lies in introducing a hybrid connection mechanism. When establishing connections, newly joining nodes do not follow a single rule but instead collaboratively perform random attachment and preferential attachment. Through theoretical derivation and simulation validation, this study demonstrates that by adjusting the cooperative weighting between these two mechanisms, the overall network degree distribution can achieve a continuous and controllable transition between a uniform distribution and a power-law distribution. The former effectively disperses attack risks and enhances robustness, while the latter facilitates the formation of hub nodes, shortening transmission paths to reduce latency. Experimental results based on the real-world road network of Beijing indicate that this model can precisely regulate node connection heterogeneity, attack resistance, and average transmission path length through the reshaping of the underlying topology. This provides a forward-looking and practical governance paradigm for constructing next-generation VANETs capable of dynamically adapting to complex environments.
\end{abstract}


\section{Introduction}
The Vehicular Ad-hoc Network (VANET) serves as the core infrastructure for Intelligent Transportation Systems (ITS) and a critical enabler for the deep integration of vehicles, roadside units, and cloud services \cite{10852542,10707134}. By establishing real-time, dynamic Vehicle-to-Vehicle (V2V) and Vehicle-to-Infrastructure (V2I) communication capabilities, it directly empowers critical applications such as real-time collision warning, dynamic traffic control, and cooperative autonomous driving \cite{RenUAV}. Consequently, VANET fundamentally determines the ultimate efficacy of ITS in enhancing road safety, traffic efficiency, and user experience \cite{RenITS,11025163,11000096}.

However, the practical deployment of this promising vision faces significant challenges. The predominant communication standard for VANETs, IEEE 802.11p, typically has a limited coverage radius of 300-500 meters, imposing an inherent constraint on the perceptual range of On-Board Unit (OBU) nodes \cite{4526014}. A more critical issue is the high mobility of vehicles (e.g., reaching 120 km/h on highways), which makes the real-time acquisition of global network topology information extremely costly, or even entirely infeasible in scenarios involving massive, densely interacting nodes \cite{RenJCICE,10582531,10982148}. These two inherent constraints collectively dictate that the topological evolution of VANETs cannot rely on global information. Instead, connections must be established based on data from neighboring nodes within the communication range, thereby giving rise to the localized VANET paradigm as a fundamental operational model.

However, while circumventing the need for global information, this localized paradigm introduces distinct governance challenges that directly impact the reliability of intelligent transportation applications. The first issue is robustness against targeted attacks. The presence of high-degree hub nodes, such as critical Roadside Units (RSUs), makes the network vulnerable. If such a node is compromised by jamming or physical destruction, it can lead to localized network failure \cite{10930944}, a risk especially critical in congested urban areas \cite{10016623}. Conversely, the second challenge concerns low-latency transmission. High-speed scenarios require millisecond-level responses for applications like collision warning. Without efficient hub nodes, information must travel through multiple hops, increasing the average path length and transmission delay beyond acceptable limits for real-time operation \cite{11148176}. The core challenge, however, lies in dynamic adaptation. Different traffic scenarios impose conflicting requirements: urban congestion demands robust, interference-tolerant topologies, while highway environments prioritize minimal latency. Existing static models cannot dynamically adjust network properties according to operational context. Relying on a single, fixed configuration in dynamic environments constitutes a major bottleneck for the practical deployment of VANETs.

The academic community has explored VANET governance challenges from multiple dimensions, which can be broadly categorized as follows. Research in topology management and optimization focuses on constructing robust and efficient network structures. Existing efforts include centralized periodic interventions for long-term routing planning at intersections \cite{10568449}, and improved multi-relay mechanisms to reduce control overhead and optimize latency and throughput \cite{Ahmed2024MultipointRP}. A systematic review of umanned aerial vehicle network topology control \cite{ALAM2022103495} also offers cross-domain algorithmic insights. Furthermore, studies have revealed the fundamental impact of topology on system resilience: while traditional nearest-neighbor topologies exhibit security flaws \cite{9678129}, k-nearest neighbor topologies achieve a better balance between performance and security. Connectivity analysis in platoon-based VANETs \cite{9173276} further quantifies performance gains from structured vehicle formations. However, most approaches focus either on optimizing specific static or quasi-static topologies, or merely revealing inherent properties of particular structures. They lack a general mechanism capable of proactively and precisely reshaping the underlying degree distribution of the network according to dynamic scenario requirements.

In the domain of routing and resource management, research focuses on efficient data dissemination over a given topology. Examples include integrating Manhattan grid topology with intelligent algorithms to enhance communication efficiency and resist jamming attacks \cite{Traffic_coordination}, dynamically allocating resources using fuzzy logic and Markov chains to optimize throughput and latency \cite{10910907,10469447}, and reducing total video delivery delay through joint topology and cache optimization \cite{10461081}. In urban settings, hybrid routing protocols incorporating mobility models have proven effective in improving delivery rate and reducing latency \cite{11089478}. While these schemes achieve effective traffic governance on existing networks, their performance ceiling is constrained by the inherent properties of the underlying topology. If the network itself is vulnerable due to overly centralized hubs, or suffers from excessively long paths due to overly uniform connectivity, upper-layer routing and resource scheduling alone can achieve only limited effectiveness.

Finally, in the domain of security, privacy, and trust governance, research aims to identify and eliminate malicious nodes to enhance network security \cite{Security}. Trust management models leveraging blockchain and machine learning \cite{10261426,9896700} can effectively assess vehicle credibility and filter misinformation, while efficient anonymous authentication schemes \cite{ZHAO2025} enable traceability of malicious vehicles without compromising privacy. A comprehensive survey on identity-based authentication \cite{10852542} systematically outlines the trade-offs in this field. These technologies are indispensable for VANET governance, but their primary role lies in filtering network participants rather than shaping the underlying connectivity structure. As revealed in \cite{10704300}, blackhole attacks can easily double latency and reduce throughput, underscoring that even within a "trusted" network, the inherent fragility of the topology itself remains a critical threat.

In summary, while existing research has made significant progress in protocol optimization, traffic management, and security authentication for VANETs, a fundamental governance challenge remains inadequately addressed: the construction of a VANET with an inherently malleable underlying topology. Such a network should dynamically and seamlessly transition between the efficiency of scale-free networks \cite{BA} and the robustness of random networks \cite{ER} in response to real-time scenario demands, prioritizing low latency on suburban highways and attack resistance in congested urban areas. Most existing models adhere to a fixed degree distribution pattern, rendering them incapable of this cross-paradigm adaptive governance. This limitation constitutes a core bottleneck hindering the transition of VANETs from theoretical frameworks to large-scale practical deployment.

To address this gap, this paper innovatively proposes a schedulable degree distribution model for localized VANET governance. The core of this model lies in introducing a uniform-preference connection probability. Through theoretical derivation and simulation, we demonstrate that the model enables a continuous, controllable transition of the network degree distribution between an attack-resistant exponential distribution and a low-latency power-law distribution. This offers a forward-looking and flexible governance paradigm for dynamic traffic environments. The core contributions are:
\begin{itemize}
	\item It systematically analyzes the degree distribution limitations in basic localized VANET models and proposes a schedulable degree distribution model. By integrating a hybrid connection mechanism that balances uniform and preferential attachment, the model achieves dynamic switching between robustness-oriented exponential and latency-oriented power-law degree distributions.
	\item Using the mean-field approach, it mathematically derives the model's degree distribution properties, explicitly quantifying the distribution patterns under different connection probabilities. Complete differential equation solving and continuous-field approximation provide a rigorous mathematical foundation, validating the model's design.
	\item An experimental environment is built using the SUMO traffic simulator with a real-world road network east of Beijing's Forbidden City. Degree distribution statistics confirm the theoretical derivation. Attack resistance is verified via targeted and random attack tests, while transmission efficiency is evaluated by measuring the average path length, collectively demonstrating the model's governance effectiveness in practical VANET scenarios.
\end{itemize}

The remainder of this paper is organized as follows. Section II analyzes the limitations of existing localized VANET models. Section III details the proposed adjustable degree distribution model. Section IV provides a theoretical derivation of the model's degree distribution properties. Section V presents the simulation setup and a quantitative analysis of the model's governance performance. Finally, Section VI concludes the paper and suggests future research directions.

\section{Analysis of Degree Distribution Limitations in Basic Localized VANET Models}
Basic localized VANET models, built upon the local-world assumption of the BA model, restrict node connections to within communication range \cite{YihanZhang}. However, their reliance on a single attachment rule results in a fixed degree distribution, lacking schedulability. This section outlines the core logic of these baseline models, followed by an analysis of their inherent limitations due to static degree distributions.

\subsection{Core Logic of the Baseline Model}
In the evolution of the BA scale-free network model, new nodes require global network information and establish connections via preferential attachment, ultimately forming a power-law degree distribution with an exponent of 3 \cite{BA,RevModPhys}. However, in VANETs, the communication range of OBUs is limited by the coverage radius of wireless technology (e.g., typically 300–500 m for IEEE 802.11p), and high vehicle mobility makes global information acquisition prohibitively costly. This necessitates the introduction of a local-world concept. The evolution of the basic localized VANET model proceeds as follows.

\textbf{Initial Network Construction.} Initially, $m_0$ core nodes, comprising $s$ RSUs fixed at intersections or road segments and $m_0-s$ OBUs, are deployed in the monitoring area. These nodes form $e_0$ initial links (e.g., full connectivity between RSUs and between RSUs and nearby OBUs) via the IEEE 802.11p protocol.

\textbf{New Node Joining and Local World Selection.} At each time step, a newly joined OBU detects all existing nodes within its communication range and randomly selects $M$ nodes as its local world. It then establishes connections with $m$ ($m \le M$) nodes from this local world.

\textbf{Local Preferential Attachment.} During network growth, each newly joined OBU bases its attachment decisions not on global information, but on its local world. This mechanism better reflects the real VANET scenario of limited communication range and localized information. Building upon the preferential attachment principle of the classic BA scale-free network model but constrained by the local world, the probability $\Pi_{\text{Local}}(k_i)$ that a new node connects to an existing node $i$ within its local world is a joint probability determined by two events:
\begin{itemize}
	\item Event 1: Node $i$ is included in the new node's local world. Assuming the total number of nodes at time step $t$ is $N(t)=m_0+t$, where $m_0$ is the initial number of nodes, the new node randomly selects $M$ nodes from its communication range to form its local world ($M\le N(t)$). Thus, the probability that any existing node $i$ is included in this local world is:
	\begin{equation}
		\Pi(i \in \text{Local}) = \frac{M}{m_0+t}.
	\end{equation}
	This term reflects the impact of global network growth dynamics on node visibility.
	\item Event 2: Node $i$ is selected via preferential attachment within the local world. Inside a given local world, the new OBU follows the rich-get-richer principle, preferring to connect to nodes with higher degrees. Let ${\textstyle \sum_{j\in{\text{Local}}}k_j}$
	be the sum of the degrees of all nodes in the local world. The conditional probability that node $i$ is selected given its inclusion in the local world is:
	\begin{equation}
		\Pi(\text{Select}\;i | i \in \text{Local})=\frac{k_i}{{\textstyle \sum_{j\in{\text{Local}}}k_j}}.
	\end{equation}
	This term ensures that connection preference is proportional to node degree within the local scope.
\end{itemize}

Integrating the above two events, the final probability for a new OBU to connect to an existing node $i$ is therefore given by the product of these two probabilities:
\begin{equation}
	\Pi _{\text{Local}}(k_i)=\Pi(i\in \text{Local})\cdot \frac{k_i}{ {\textstyle \sum_{j\in{\text{Local}}}k_j} }=\frac{M}{m_0+t}\cdot \frac{k_i}{{\textstyle \sum_{j\in{\text{Local}}}k_j} }.
\end{equation}

\subsection{Scenario Mapping of Degree Distribution Limitations}
The degree distribution of the baseline model is solely determined by the local world size $M$, constrained by $m \leq M \leq m_0 + t$. Three typical scenarios are analyzed based on the value of $M$:

\textbf{(1) When $M=m$:} The number of nodes in the new OBU's local world exactly equals its required number of connections $m$. Consequently, the new node must establish links with all nodes in its local world. In this scenario, network evolution involves only a growth mechanism without preferential attachment. This resembles a blind connection scenario in VANETs, such as when a vehicle rapidly enters a suburban or remote highway area without RSU coverage, forcing it to form temporary connections with a limited number of nearby vehicles without the ability to select highly connected nodes.

Based on the mean-field approach, the discrete evolution of node degree is approximated using a continuous real variable. The change in degree $k_i$ of node $i$ with respect to time step $t$ can be described by the following equation:
\begin{equation}
	\frac{\partial k_i}{\partial t}=\frac{m}{m_0+t}  
\end{equation}
Solving this differential equation yields an exponential degree distribution of the form:
\begin{equation}
	P(k) \propto e^{-\frac{k}{m}}
\end{equation}
The resulting uniform degree distribution, lacking prominent hubs, disperses the impact of targeted attacks (e.g., jamming, physical destruction), thereby preserving basic connectivity even under partial node failures. However, the absence of hub nodes for efficient multi-hop forwarding increases the average path length, ultimately failing to meet the low-latency demands of VANET applications such as collision warning and real-time coordination.

\textbf{(2) When $M=m_0+t$:} The new OBU's local world encompasses all nodes in the entire network, reducing the model to the BA scale-free model. In this scenario, new nodes acquire global network information and select connections based on preferential attachment. This is only feasible in small-scale, low-mobility environments like vehicle networks in confined areas, as global information perception is impractical for large-scale VANETs. The probability of an existing node $i$ being selected by a new node is:
\begin{equation}
	\Pi (k_i)=\frac{k_i}{ {\textstyle \sum_{j}k_j} }, 
\end{equation}
Each newly joined OBU establishes $m$ new connections. Thus, at time step $t$, the total number of edges is $E=t\cdot m$. Since each edge contributes 2 to the total degree count, the sum of all node degrees in the network is 
\begin{equation}
	{\textstyle \sum_{j}k_j} = 2E =2tm.
\end{equation}
Under the preferential attachment mechanism in the BA model, the probability that a new node selects an existing node $i$ is given by:
\begin{equation}
	\Pi (k_i)=\frac{k_i}{ {\textstyle \sum_{j}k_j} }=\frac{k_i}{2tm}
\end{equation}
Assuming one new OBU joins per time step and establishes $m$ new connections, the rate of change of node $i$'s degree equals the probability of its selection multiplied by $m$, expressed as:
\begin{equation}
	\frac{\partial k_i}{\partial t} =m \cdot \Pi(k_i)
\end{equation}
Substitution yields the temporal evolution of node $i$'s degree $k_i$ with respect to time step $t$ as:
\begin{equation}
	\frac{\partial k_i}{\partial t} = m \cdot \frac{k_i}{2tm} = \frac{k_i}{2t}
\end{equation}

This demonstrates that a higher degree $k_i$ leads to a greater rate of change $\frac{\partial k_i}{\partial t}$, reflecting the rich-get-richer nature of preferential attachment. Meanwhile, as the network matures (larger $t$), the impact of new connections on individual node degrees dilutes. Solving the equation confirms the degree distribution follows a power law, expressed as:
\begin{equation}
	P(k) \propto 2m^2 k^{-3}
\end{equation}
This configuration generates a small number of highly connected hubs (e.g., critical RSUs or highly active vehicles), which facilitate efficient data forwarding. However, this makes the network highly vulnerable to targeted attacks, as the compromise of a single hub can trigger localized network paralysis.

\textbf{(3) When $m<M<m_0+t$:} This represents the typical operational regime of the localized VANET model. New OBUs select their local world from a partial set of neighbors within communication range, adhering to practical constraints while enabling emergent scale-free characteristics through localized preferential attachment. The degree $k_i$ of node $i$ evolves with time step $t$ as governed by:
\begin{equation}
	\frac{\partial k_i}{\partial t} = m \cdot \frac{M}{m_0 + t} \cdot \frac{k_i}{\sum_{j \in \text{Local}} k_j}
\end{equation}

The total number of links in the network is the sum of the initial links $e_0$ and the $m$ links added per time step over $t$ steps, giving $mt + e_0$. With a total of $t + m_0$ nodes, and since each link contributes 2 to the total degree, the average node degree $\langle k \rangle$ at time step $t$ is:
\begin{equation}
	\left \langle k \right \rangle = \frac{2(mt + e_0)}{m_0 + t}
\end{equation}
The sum of the degrees of nodes within the local world equals the average degree multiplied by $M$:
\begin{equation}
	\sum_{j \in \text{Local}} k_j = M \cdot <k> = \frac{2M(mt + e_0)}{m_0 + t}
\end{equation}
Integrating these expressions, the resulting degree distribution for this scenario is derived as:
\begin{equation}
	P(k) \approx 2m^2 k^{-3}
\end{equation}
This network exhibits both local constraints and preferential attachment, producing a degree distribution that approximates a power law, similar to a scale-free network but distinct from the pure BA model due to its localized preference, demonstrating an extension of the BA model that preserves scale-free properties under local constraints.

Analysis of the three scenarios reveals that as the local world size $M$ increases, the network's degree distribution transitions from exponential to power-law, shifting its properties from attack resistance towards low-latency transmission. However, the model inherently lacks dynamic schedulability—it cannot actively balance robustness and transmission efficiency according to specific traffic scenarios (e.g., prioritizing low latency on highways). This highlights the critical need for a VANET model capable of achieving on-demand degree distribution scheduling through active parameter control.

\section{Design of an Adjustable Distribution-Based Localized VANET Model}
To address the limitations of the baseline model, this paper introduces a hybrid connection mechanism, enabling continuous degree distribution scheduling between exponential and power-law regimes. This section details the model's design principles, evolutionary steps, and core mechanism.

\subsection{Design Principles and Core Logic for Degree Distribution Scheduling}
As a dynamic complex network, VANET topology evolution must satisfy the following core requirements, which form the basis for our model enhancement:
\begin{itemize}
	\item Distributed Construction: Due to mobility and limited communication range, nodes cannot access global network information. The model must support distributed networking using local data, avoiding the high overhead of centralized control, while accommodating different node types.
	\item Dynamic Robustness Adaptation: VANETs face both random and targeted attacks. The fixed degree distribution of the baseline model cannot simultaneously address both threats, necessitating a tunable attachment rule to dynamically adapt robustness.
	\item Low-Latency Transmission: Core VANET applications require minimal latency. The ultra-small-world characteristic of scale-free networks, exhibiting short average path lengths, is crucial for meeting this requirement. Model improvements must strive to preserve low-latency characteristics while adjusting robustness.
\end{itemize}

To address these requirements, we introduce a hybrid attachment probability $p$ to enable dynamic degree distribution scheduling. Each new OBU establishes connections via uniform attachment with probability $p$, and preferential attachment with probability $1-p$. By adjusting $p$, the degree distribution can be dynamically tuned between an exponential distribution (for attack resistance) and a power-law distribution (for low latency), precisely adapting to different scenarios. When $p=1$, uniform attachment dominates, forming an exponential degree distribution (interference-resistant mode). When $p=0$, preferential attachment prevails, yielding a power-law distribution (low-latency mode). When $0<p<1$, hybrid attachment produces a tunable power-law variant.

\begin{figure}[htbp]
	\centerline{\includegraphics[width=3.5in,keepaspectratio]{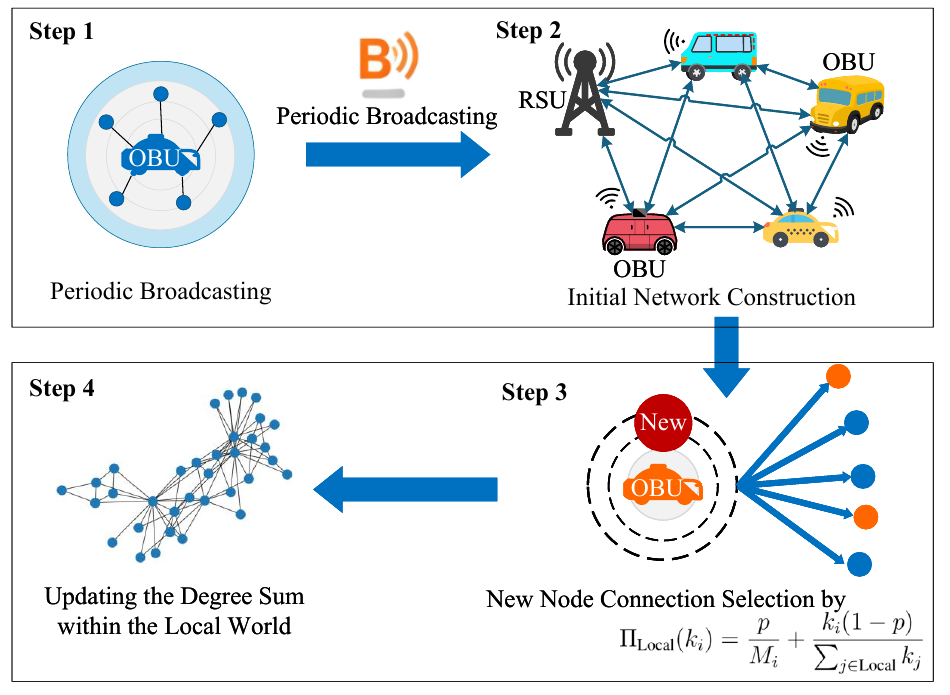}}
	\caption{Process diagram for model evolution steps.}
	\label{fig1}
\end{figure}

\subsection{Model Evolution Process}
The model evolution, illustrated in Fig. 1, adheres to VANET deployment scenarios and protocol specifications through the following steps:

\textbf{Step 1: Local World Sensing.} Nodes periodically broadcast IEEE 802.11p Beacon frames, recording neighbor IDs within communication range to define their local world.

\textbf{Step 2: Initial Network Construction.} Initialize the network with $s$ RSUs and $m_0 - s$ OBUs, forming a fully connected topology with $e_0 = \frac{m_0(m_0 + 1)}{2}$ initial links.

\textbf{Step 3: New OBU Joining and Connection.} At each time step, $n$ new OBUs join. If a new OBU's local world contains at least $m$ nodes, it selects $m$ nodes using the proposed probabilistic model; otherwise, it connects to all existing nodes in its local world.

The core innovation lies in the hybrid attachment probability $p$, mathematically expressed as a weighted combination of uniform and preferential attachment. The connection probability for node $i$ in the local world of the new node is defined as:
\begin{equation}
	\Pi_{\text{Local}}(k_i) = \frac{p}{M_i} + \frac{k_i(1 - p)}{\sum_{j \in \text{Local}} k_j}
\end{equation}
where $k_i$ denotes the degree of node $i$, $\sum_{j \in \text{Local}} k_j$ represents the total degree of nodes within the local world, and $M_i$ is the total number of nodes in the new node's local world. 

Note that Equation (16) is a theoretical probability expression. Two key issues in practical simulations require engineering solutions: 1) Floating-point precision limitations can lead to some nodes’ calculated probabilities approaching 0 (not strictly 0) or the total probability deviating from 1, violating random sampling’s probability axioms; 2) In extreme scenarios (e.g., high proportion of low-degree nodes in the local world), valid selectable nodes may be fewer than the preset connection count m, causing direct probability-based sampling to fail. To ensure connection selection’s stability, legality, and accuracy, final node sampling follows two steps:
\begin{itemize}
	\item Probability Preprocessing and Normalization: Extract candidate nodes and their initial probabilities from the local world, and construct a probability array. Replace extremely small values (less than $\epsilon$) with \(\epsilon\) to avoid effective probability loss from rounding errors, then normalize the array to ensure the total probability is strictly 1, meeting random sampling’s mathematical premise.
	\item Without-Replacement Sampling: Count valid nodes (probability > $\epsilon$). If fewer than $m$, adopt random without-replacement sampling as a fallback; otherwise, sample based on normalized hybrid attachment probabilities. This adheres to Equation (16)’s design, avoids duplicate connections to the same node, and aligns with VANETs’ single-link communication.
\end{itemize}

These engineering adaptations realize seamless integration of the theoretical model and simulation practice, ensuring the hybrid attachment strategy aligns with the original design and boosting the algorithm’s robustness in complex scenarios.

\textbf{Step 4: Real-time Degree Distribution Update.} New OBUs broadcast updated degree information via extended Beacon frames (2-byte payload expansion). Neighbors update local degree sums $\sum_{j \in Local} k_j$ to enable dynamic feedback regulation.

The following section will demonstrate that continuous adjustment of $p \in [0,1]$ enables smooth transition of attachment behavior from purely uniform to purely preferential, thereby providing rigorous mathematical support for dynamic scheduling of VANET degree distributions across exponential and power-law regimes.

\section{Analysis of Degree Distribution Scheduling Characteristics}
This section employs the mean-field approach to theoretically derive the degree distribution under different values of $p$, clarifying its schedulable range and parametric behavior to validate the model's adaptability across VANET scenarios.

\noindent\textbf{(1) When \( p = 1 \): Purely Uniform Attachment}

A new OBU selects nodes within its local world with equal probability, reducing network evolution to a growth process without preferential attachment. The rate of change of degree  $k_i$ for node $i$ is:
\begin{equation}
	\frac{\partial k_i}{\partial t} = \frac{M_i}{m_0 + t} \cdot \frac{m}{M_i} = \frac{m}{m_0 + t}
\end{equation}
Given the initial condition at its joining time \( t_i \): $k_i(t_i) = m$. Integrating the rate equation yields:
\begin{equation}
	k_i(t) = m \int_{t_i}^{t} \frac{1}{m_0+\tau}  d\tau + C
\end{equation}
Solving with the initial condition gives the degree-time relationship:
\begin{equation}
	k_i(t) = m \left( \ln \frac{m_0 + t}{m_0 + t_i} + 1 \right)
\end{equation}
Under the continuous approximation, the degree distribution can be derived from the uniform joining time distribution of nodes, \( P(t_i) = 1/t \) for \( 0 \le t_i \le t \). Let \( T = m_0 + t \) and \( T_i = m_0 + t_i \), then:
\begin{equation}
	k = m \left( \ln \frac{T}{T_i} + 1 \right)
\end{equation}
which gives $T_i = T e^{1 - k/m}$.

The cumulative distribution function is given by $F(k) = P(k_i(t) \le k)$. Since $k_i(t)$ decreases as \(T_i\) increases:
\begin{equation}
	k_i(t) \le k \Leftrightarrow T_i \ge T e^{1 - k/m}
\end{equation}
With $T_i$ uniformly distributed over $[m_0, T]$ with density $\frac{1}{T - m_0}$:
\begin{equation}
	F(k) = P(T_i \ge T e^{1 - k/m}) = \frac{T - T e^{1 - k/m}}{T - m_0}
\end{equation}
For large $t$ where $T \approx t$ and $T - m_0 \approx t$: $
F(k) \approx 1 - e^{1 - k/m}$. The probability density function is then:
\begin{equation}
	P(k) = \frac{dF(k)}{dk} = \frac{e}{m} e^{-k/m}.
\end{equation}
This confirms an exponential degree distribution with uniform node degrees and no significant hubs. As \(k\) increases, \(P(k)\) decays exponentially, making high-degree nodes rare. Suitable for interference-resistant VANET scenarios (e.g., urban congestion), this distribution maintains basic connectivity under node failures with strong robustness against targeted attacks, albeit at the cost of longer average transmission paths.

\noindent\textbf{(2) When $p = 0$: Purely Preferential Attachment}

New OBUs select nodes solely through preferential attachment within their local world, reducing the model to the baseline localized VANET model. The rate of change of node $i$'s degree is:
\begin{equation}
	\frac{\partial k_i}{\partial t} = m \cdot \frac{M_i}{m_0 + t} \cdot \frac{k_i}{\sum_{j \in \text{Local}} k_j}
\end{equation}
The average node degree and local world degree sum are:
\begin{equation}
	\langle k \rangle = \frac{2mt + m_0(m_0 - 1)}{m_0 + t},
\end{equation}
\begin{equation}
	\sum_{j \in \text{Local}} k_j = M_i \cdot \langle k \rangle
\end{equation}
For large \( t \), the term \( m_0(m_0 - 1) \) becomes negligible, yielding:
\begin{equation}
	\langle k \rangle \approx \frac{2mt}{m_0 + t}
\end{equation}
Substituting into the rate equation gives:
\begin{equation}
	\frac{\partial k_i}{\partial t} = \frac{m}{m_0+t} \cdot \frac{k_i}{\left \langle k \right \rangle}
\approx \frac{m}{m_0+t} \cdot \frac{k_i}{\frac{2mt}{m_0+t}}
= \frac{k_i}{2t}
\end{equation}

Similarly, applying the initial condition $k_i(t_i) = m$, the degree-time relationship is solved as:
\begin{equation}
	k_i(t) = m \left( \frac{t}{t_i} \right)^{1/2}
\end{equation}

Using the continuum approach with $P(t_i) = 1/t$ for $0 \le t_i \le t$, equation (29) yields $t_i = m^2 t k^{-2}$. Since $k$ decreases with increasing $t_i$, that is, nodes that join earlier have larger degrees, the cumulative distribution function becomes:
\begin{equation}
	F(k) = P(k_i \le k) = P(t_i \ge m^2 t k^{-2}) = 1 - m^2 k^{-2}
\end{equation}
Here, it is required that \( m^2 t k^{-2} \geq t \), which is equivalent to \( k \geq m \), and \( m^2 t k^{-2} \geq 0 \) holds automatically, then:
\begin{equation}
	P(k) = \frac{dF(k)}{dk} = 2m^2 k^{-3}
\end{equation}
Thus, the degree distribution follows a power law:
\begin{equation}
	P(k) \propto 2m^2 k^{-3}
\end{equation}
This regime produces a scale-free network with few high-degree hubs, suitable for low-latency VANET scenarios (e.g., highways). Hub nodes shorten communication paths, meeting delay-sensitive application requirements, but create vulnerability to targeted attacks where hub failure may cause localized network disruption.

\noindent \textbf{(3) When $0 < p < 1$: Hybrid Attachment}

New OBUs employ both uniform and preferential attachment, representing the model's typical operational regime. For large $t$, the rate of change of node $i$'s degree is:
\begin{equation}
	\frac{\partial k_i}{\partial t} = \frac{mp}{m_0 + t} + \frac{k_i(1 - p)}{2t} \approx \frac{mp}{t} + \frac{k_i(1 - p)}{2t}
\end{equation}
Similarly, by solving the differential equation in Formula (33) and combining it with the initial condition $k_i(t_i) = m$, the relationship between the node degree and the time step can be derived, that is
\begin{equation}
	k_i(t) = \frac{m(1 + p)}{1 - p} \cdot \left( \frac{t}{t_i} \right)^{\frac{1 - p}{2}} - \frac{2mp}{1 - p}
\end{equation}

Let $\beta = \frac{1 - p}{2}$, $A = \frac{m(1 + p)}{1 - p}$, $B = \frac{2mp}{1 - p}$. Through algebraic derivation:
\begin{equation}
	k = A\left( \frac{t}{t_i} \right)^\beta - B
\end{equation}
\begin{equation}
	t_i = t \left( \frac{k + B}{A} \right)^{-1/\beta}
\end{equation}
The cumulative distribution function is:
\begin{equation}
	F(k) = P(k_i \leq k) = P\left( t_i \geq t \left( \frac{k + B}{A} \right)^{-1/\beta} \right)
\end{equation}
Since \(k_i\) decreases with increasing $t_i$, further derivation yields:
\begin{equation}
	F(k) = 1 - \frac{t \cdot \left( \frac{k + B}{A} \right)^{-1/\beta}}{t} = 1 - \left( \frac{k + B}{A} \right)^{-1/\beta}
\end{equation}

The probability density function is obtained by differentiation:
\begin{equation}
	P(k) = \frac{dF(k)}{dk} = -\frac{d}{dk} \left[ \left( \frac{A}{k + B} \right)^{1/\beta} \right]
\end{equation}
Differentiating $\left( \frac{A}{k + B} \right)^{1/\beta}$:
\begin{align*}
	\frac{d}{dk} \left( \frac{A}{k + B} \right)^{1/\beta} &= \frac{1}{\beta} \left( \frac{A}{k + B} \right)^{\frac{1}{\beta} - 1} \cdot A \cdot \left( -\frac{1}{(k + B)^2} \right) \\
	&= -\frac{1}{\beta} A^{1/\beta} (k + B)^{-1/\beta - 1}
\end{align*}
Thus:
\begin{equation}
	P(k) = -\left[ -\frac{1}{\beta} A^{1/\beta} (k + B)^{-1/\beta - 1} \right] = \frac{1}{\beta} A^{1/\beta} (k + B)^{-1/\beta - 1}
\end{equation}
Substituting \(A\), \(\beta\), and \(B\) yields the final probability density function:
\begin{equation}
	P(k) = \frac{2}{1 - p} \left[ \frac{m(1 + p)}{1 - p} \right]^{\frac{2}{1 - p}} \left[ k + \frac{2mp}{1 - p} \right]^{-\frac{3 - p}{1 - p}}
\end{equation}
Letting \(C = \frac{2}{1-p}\left [ \frac{m(1+p)}{1-p} \right ]^{\frac{2}{1-p}}\), \(a = \frac{2mp}{1-p}\), \(\gamma = \frac{3-p}{1-p}\), we obtain:
\begin{equation}
	P(k) = C \cdot (k + a)^{-\gamma} \propto (k + a)^{-\gamma}
\end{equation}
This indicates the degree distribution approximately follows a tunable power law. With \(0 < p < 1\), \(\gamma > 3\), and increasing with \(p\), the distribution transitions gradually from a steep power-law (close to BA model) to a flatter exponential form. The number of hub nodes decreases while node degrees become more uniformly distributed.

\begin{figure}[t]
	\centerline{\includegraphics[width=3in,keepaspectratio]{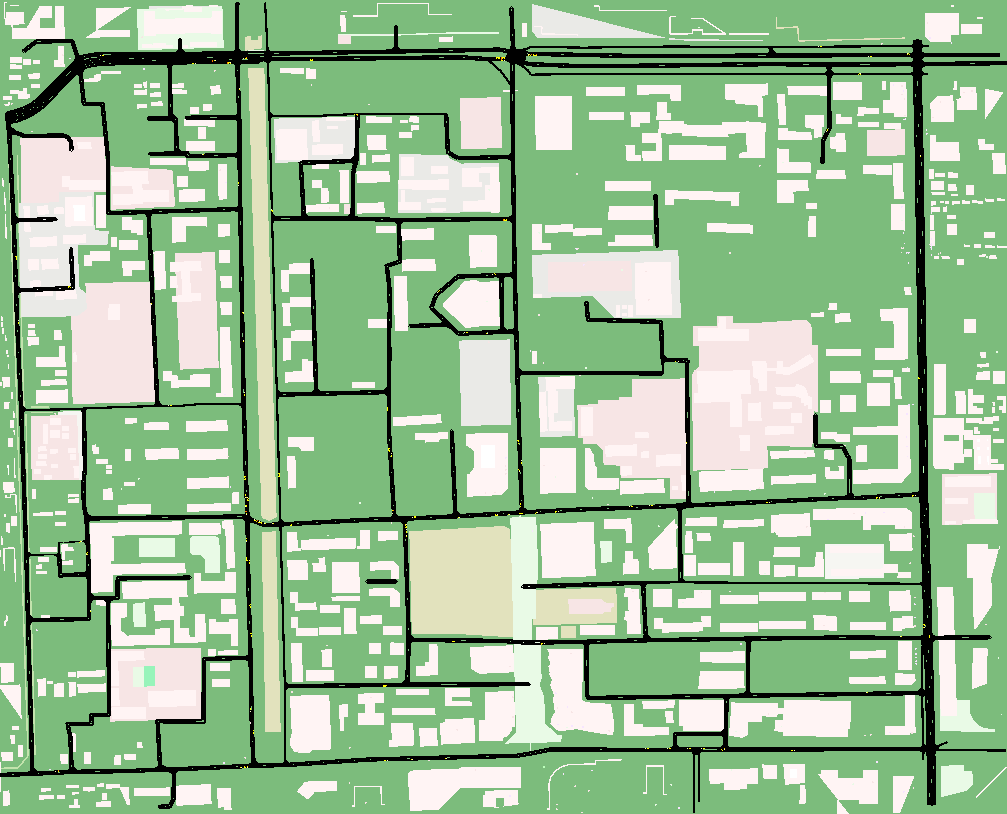}}
	\caption{Net of an area to the east of the Forbidden City in Beijing.}
	\label{fig2}
\end{figure}

\begin{algorithm}[!h]
	\caption{VANET Simulation Procedure}
	\label{alg:AOS}
	\renewcommand{\algorithmicrequire}{\textbf{Initialization:}}
	\renewcommand{\algorithmicensure}{Launch SUMO and deploy RSUs.}
	\begin{algorithmic}[1]
		\REQUIRE Configure the SUMO environment, the random seed, define data structures for storing connection pairs and node degrees.
		\ENSURE 
		\FOR{each simulation step}
		\STATE a. Advance the simulation by one step; retrieve currently active OBUs and their positions.
		\STATE b. Remove invalid connections: Disconnect nodes that are out of communication range or have left the scene; update node degrees accordingly.
		\STATE c. Establish new connections:
		\STATE  \hspace{5mm} For each OBU, filter RSUs and OBUs within communication range to form a local node set.
		\STATE   \hspace{5mm} Select $m$ nodes from local set by Equation (16).
		\STATE  \hspace{5mm} Establish new connections and update the degrees of corresponding nodes.
		\STATE d. Calculate metrics such as the proportion of the largest connected component and degree variance; record results into statistics.
		\ENDFOR
		\STATE Terminate the simulation; save every step statistics, final topology, and degree distribution to files.
	\end{algorithmic}
\end{algorithm}

\section{Simulation Experiments and Result Analysis}
To validate the degree distribution schedulability and scenario adaptability of the proposed model, an experimental environment is constructed based on SUMO and Python. All experiments are conducted on a system equipped with a 12th Gen Intel(R) Core(TM) i5-12400F CPU @ 2.50GHz processor and 16GB RAM. The software configuration consisted of Anaconda3-2021.11-Windows-x86\_64 and Python 3.8.

Using the OSMWebWizard.py tool provided by SUMO, a real road network section (1426.16 $\times$ 1145.47 m) east of Beijing's Forbidden City is extracted from OpenStreetMap as the simulation scenario. The network is processed by retaining only motor-vehicle routes and rendered as shown in Fig. 2. Vehicle traffic trajectories are generated via randomTrips.py, with one new OBU joining per second. For simplicity, RSUs are deployed at key intersections. Following the IEEE 802.11p protocol, the communication range is set to 300 m for OBUs and 500 m for RSUs. The procedure for the simulation implementation code is presented in Algorithm 1.

\begin{figure}[t]
	%\begin{minipage}{0.48\linewidth}中的{0.48\linewidth} 是调整图片列的宽度，\linewidth之前的0.48是可以调整的    
	\begin{minipage}{0.48\linewidth}
		%\vspace{3pt}  是调图片之间的间隔
		\centerline{\includegraphics[width=\textwidth]{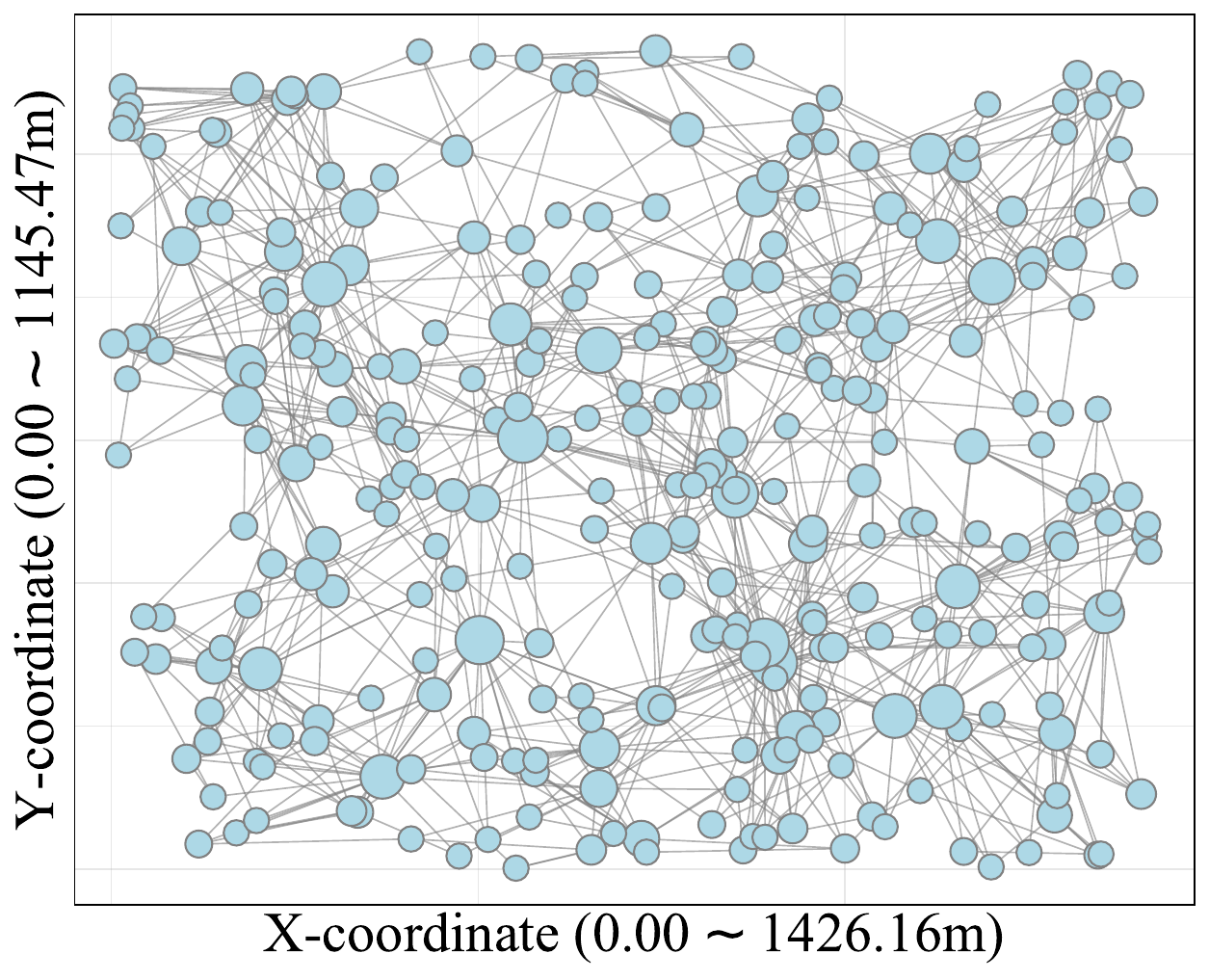}}
		
		\centerline{\includegraphics[width=\textwidth]{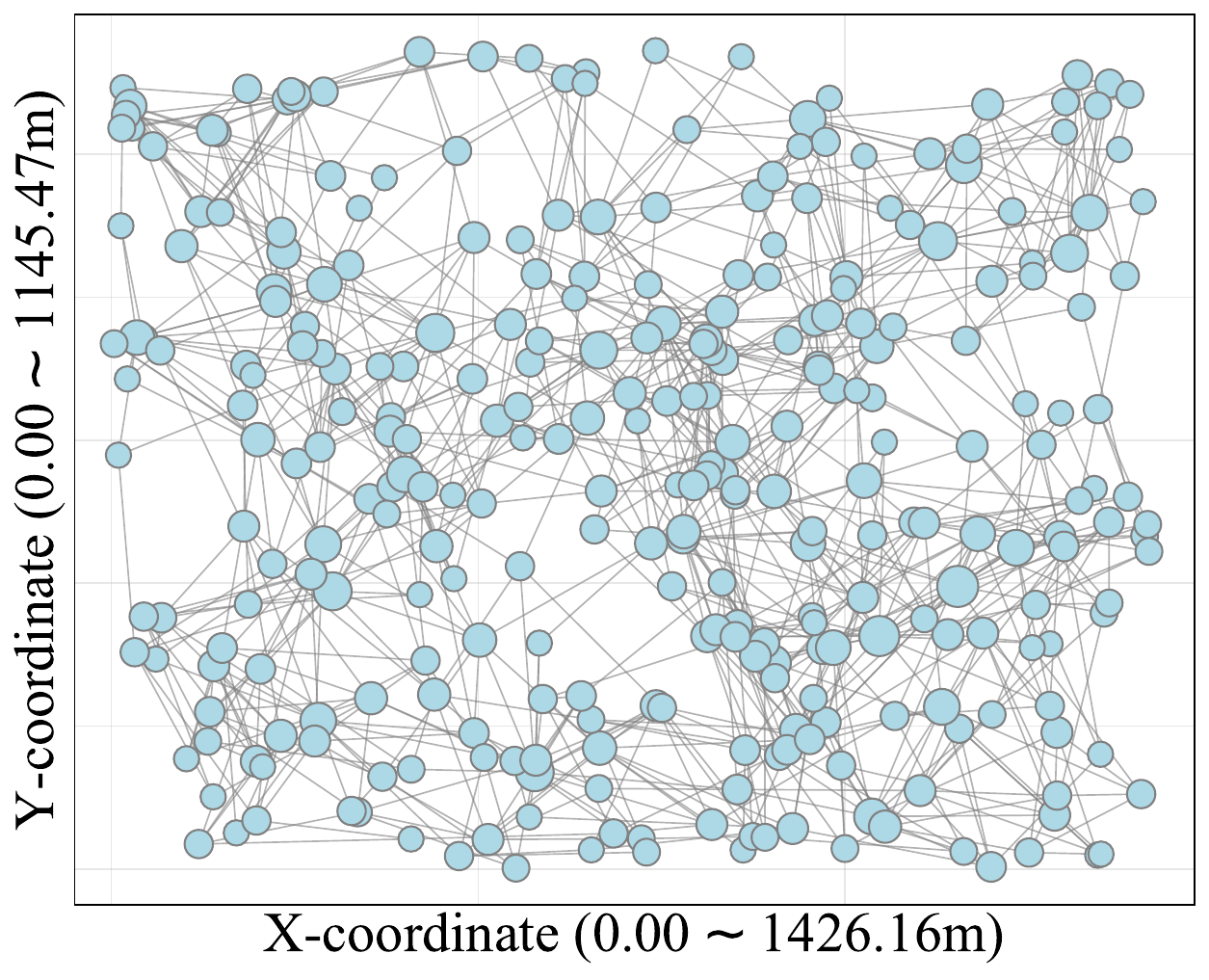}}
		
		\centerline{\includegraphics[width=\textwidth]{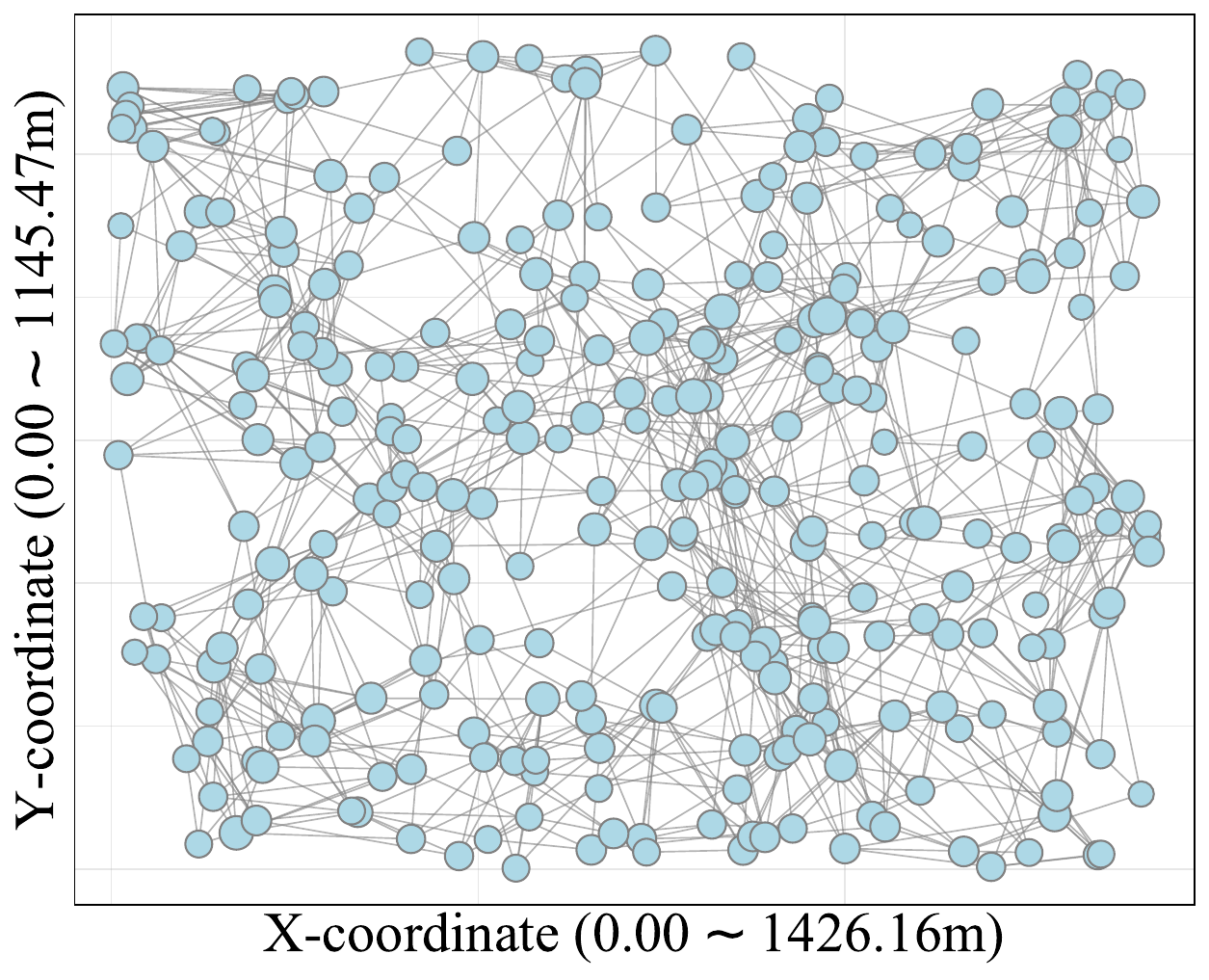}}
	\end{minipage}
	%一个 \begin{minipage}{0.48\linewidth} 就是一列
	\begin{minipage}{0.48\linewidth}
		\centerline{\includegraphics[width=\textwidth]{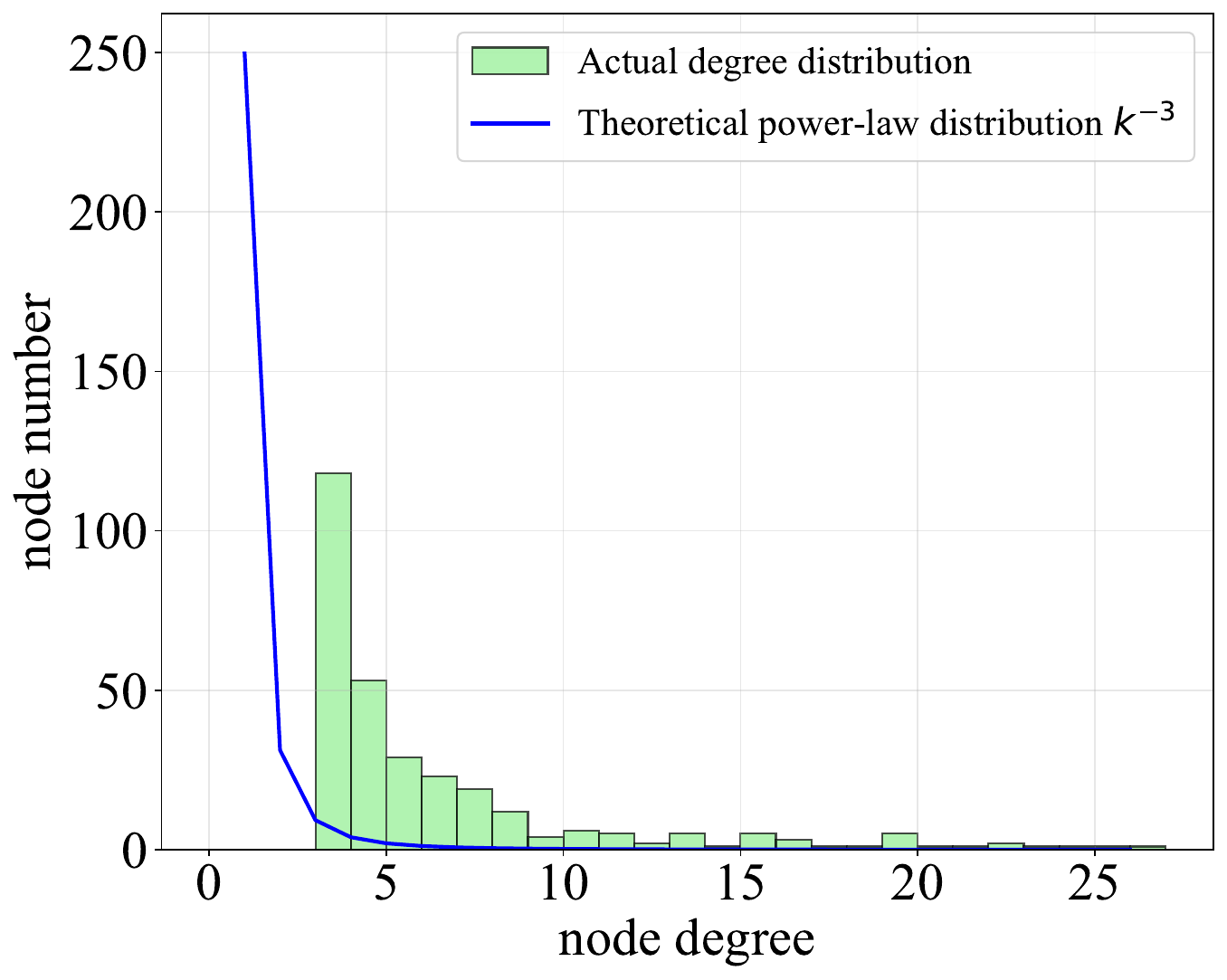}}
			
		\centerline{\includegraphics[width=\textwidth]{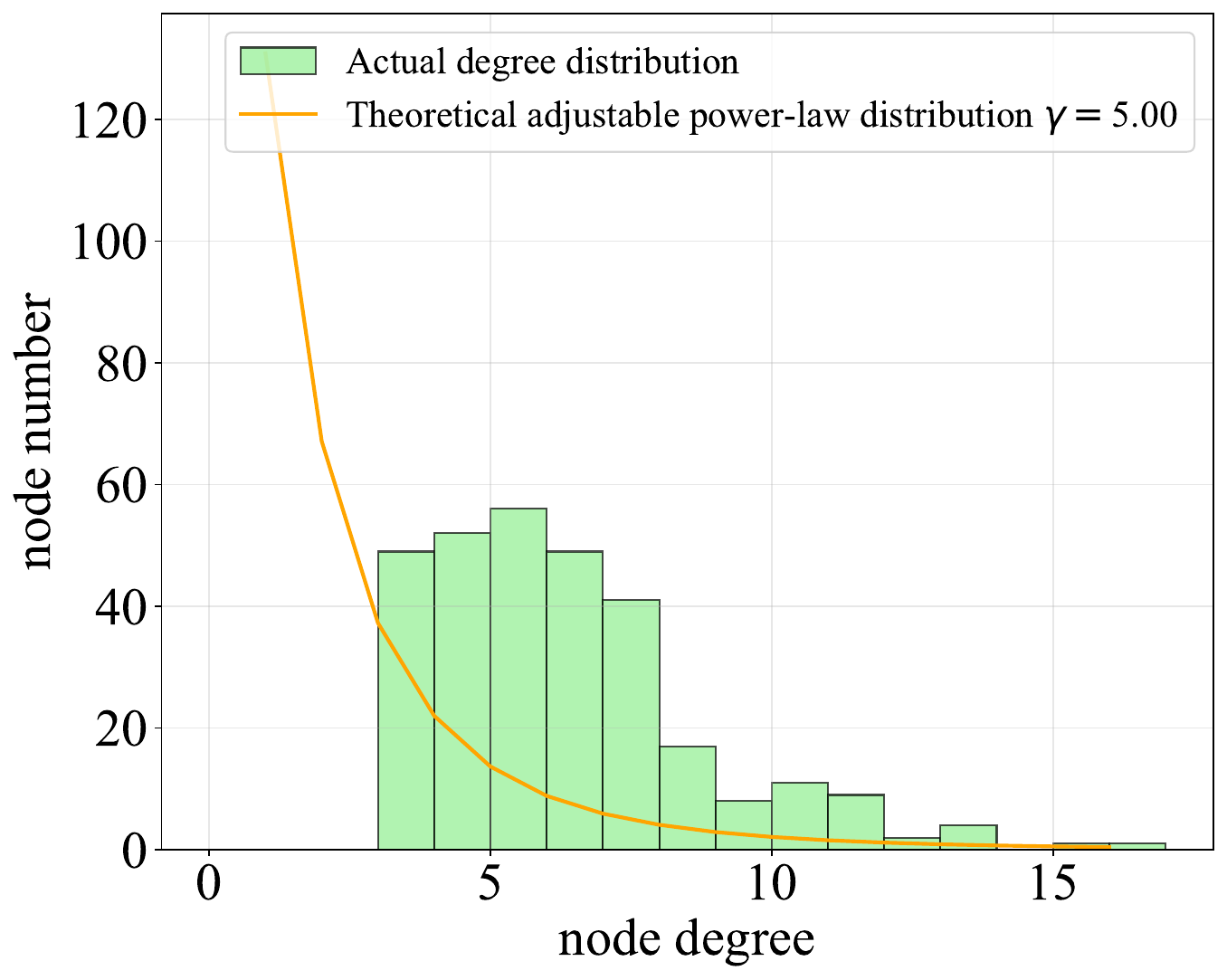}}
			
		\centerline{\includegraphics[width=\textwidth]{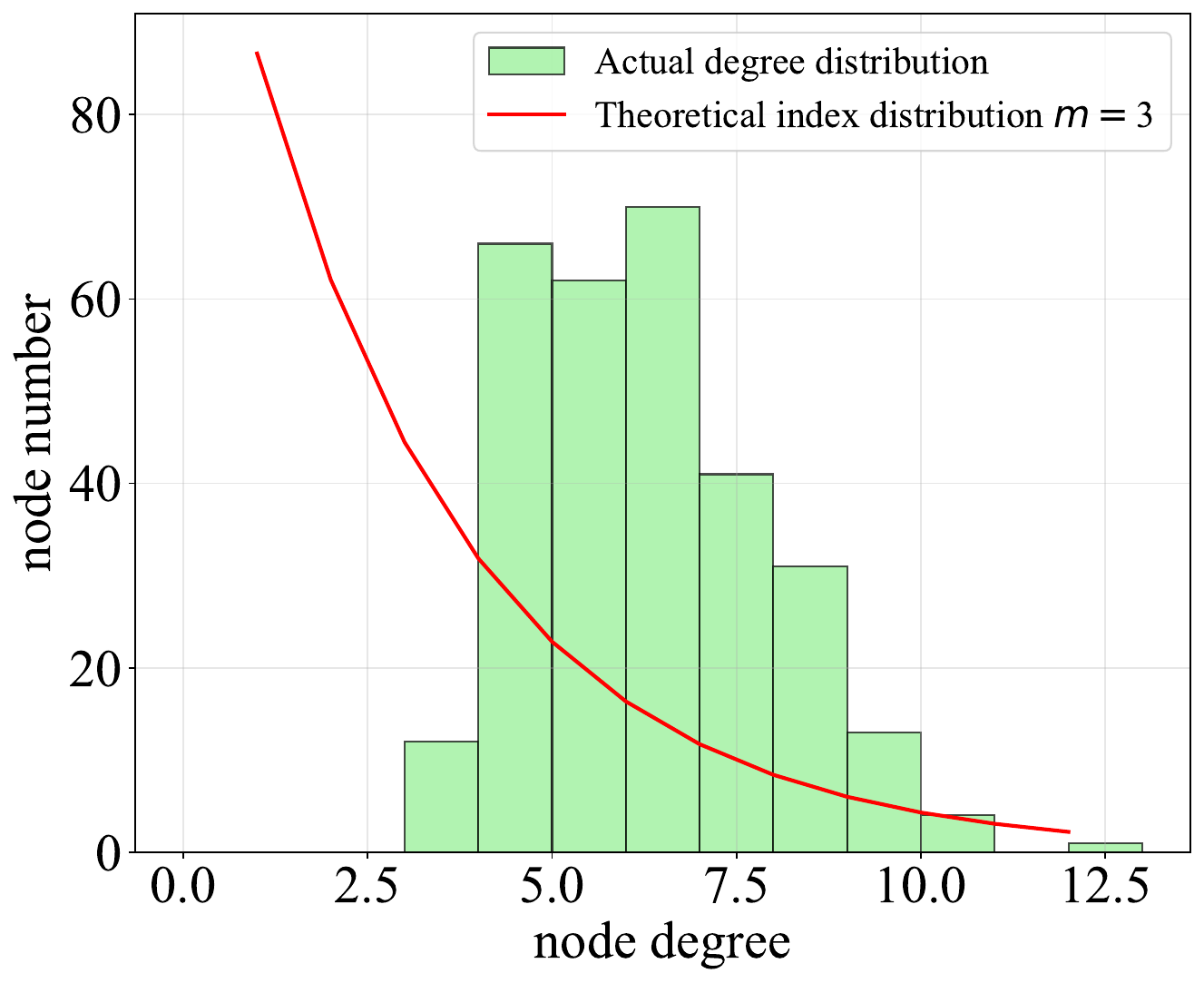}}
	\end{minipage}
	\caption{Topological links and corresponding degree distributions under different $p$, with upper, middle, and lower subfigures corresponding to $p = 0.0$, $0.5$, and $1.0$, respectively.}
	\label{fig3}
\end{figure}

To clearly observe the distribution patterns under different values of $p$, we randomly generated 200 OBU nodes within the imported road network and constructed topological connections according to different $p$ values. The resulting degree distributions are shown in Fig. 3. The experimental results align well with theoretical expectations. In the topology diagrams, node radius scales with degree to visually emphasize hubs. When $p = 0.0$, distinct large hub nodes emerge, forming a centralized rich-get-richer structure. This configuration shortens transmission paths in highway scenarios but remains vulnerable to hub failures. At $p = 0.5$, hub concentration moderates, balancing latency optimization and interference resistance for suburban environments. With $p = 1.0$, nodes exhibit minimal size variation with uniformly distributed links, enhancing resilience against node failures in urban congestion. 

In terms of degree distribution, the empirical distribution at $p = 0.0$ closely aligns with the theoretical power-law $k^{-3}$, exhibiting characteristics of a scale-free network where a few nodes possess extremely high degrees while the majority have low connectivity. At $p = 0.5$, the empirical distribution fits well with the theoretical tunable power-law with $\gamma = 5.0$, demonstrating a degree inequality that lies between the cases of $p = 0$ and $p = 1$. When $p = 1.0$, the empirical distribution follows the theoretical exponential trend. Due to the pure uniform attachment rule, node degrees concentrate around the mean, eliminating the prevalence of very low-degree nodes. This results in a small first-bin value, a typical feature distinguishing exponential from power-law distributions, reflecting how uniform attachment leads to degree distributions clustered near the average.

\begin{figure}[t]
	\centerline{\includegraphics[width=3.2in,keepaspectratio]{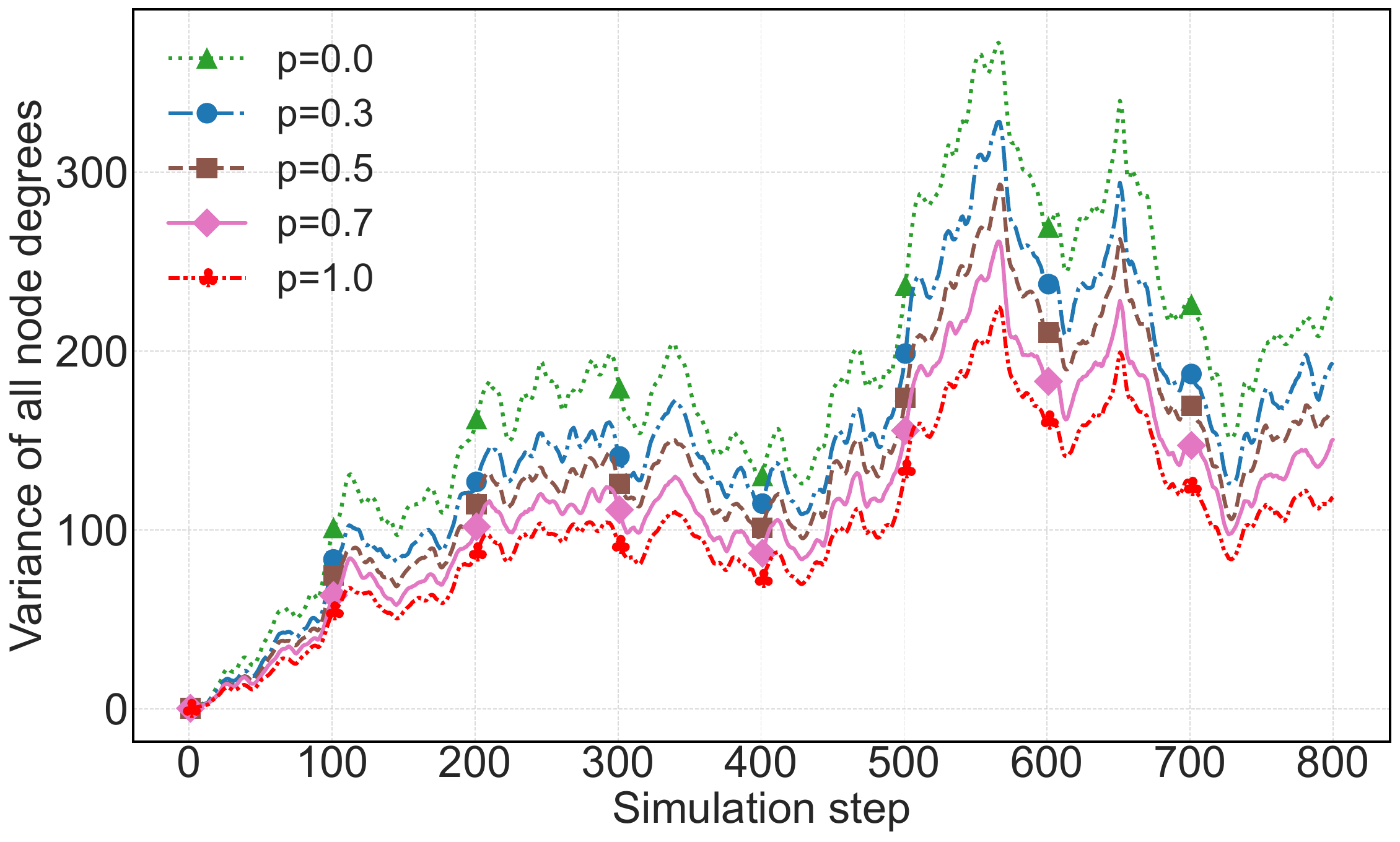}}
	\caption{Variance of all node degrees.}
	\label{fig4}
\end{figure}

\begin{figure}[t]
	\centerline{\includegraphics[width=3.2in,keepaspectratio]{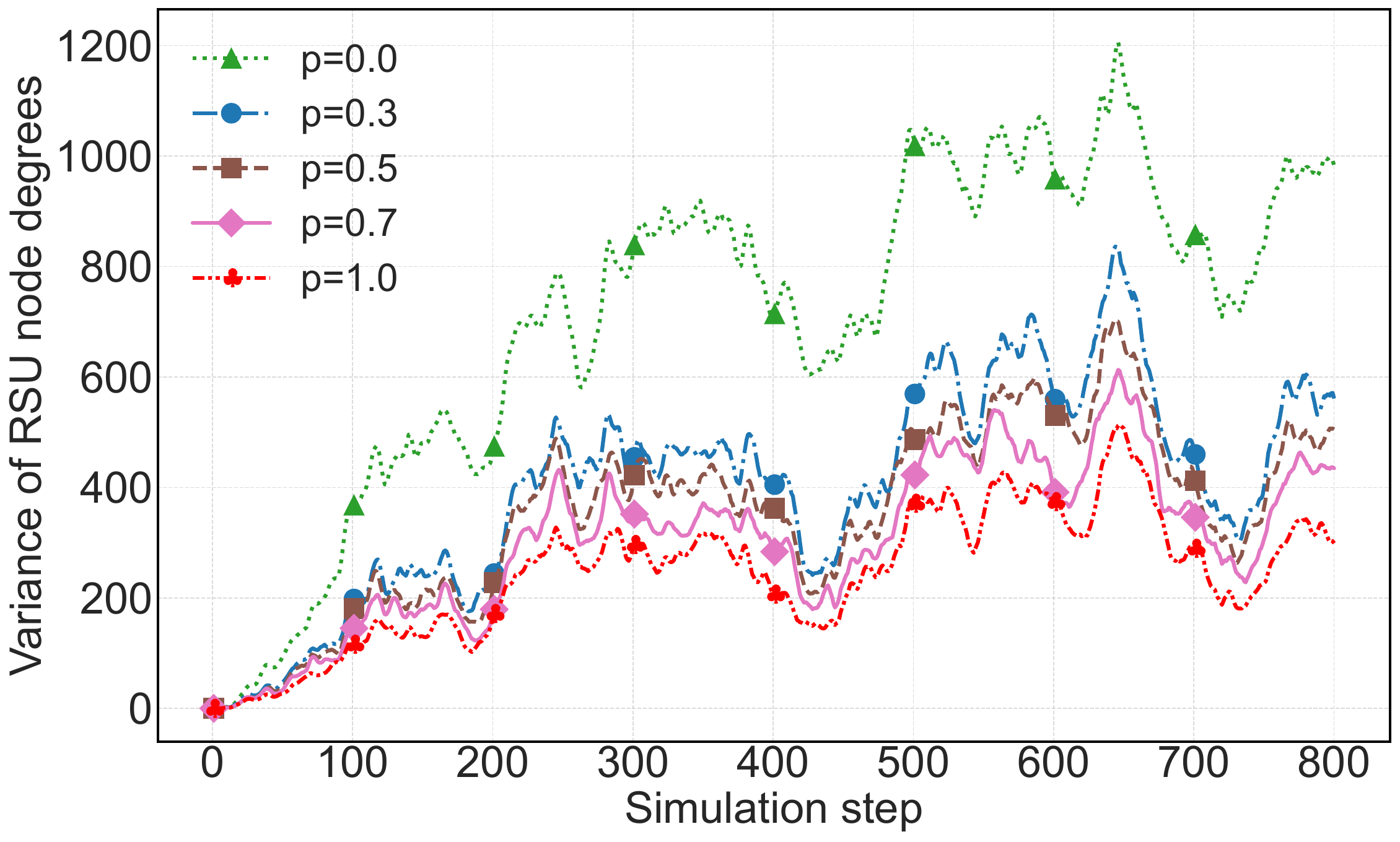}}
	\caption{Variance of RSU node degrees.}
	\label{fig5}
\end{figure}

Figs. 4 and 5 illustrate the dynamic evolution of node degree variance under different hybrid attachment probabilities $p$ over 800 simulation steps. Throughout the simulation, the degree variance exhibits significant fluctuations due to continuous topological dynamics. The initial increase arises from new OBUs joining and establishing connections, rapidly amplifying degree disparities during the early growth phase. RSUs demonstrate a more prolonged growth period, as their larger communication range and fixed positioning enable sustained link formation with incoming nodes. 
Notably, degree variance consistently decreases with increasing $p$ for both all nodes and RSUs. This trend confirms that higher uniform attachment probability effectively suppresses degree heterogeneity by distributing connections more evenly within local worlds, thereby mitigating the rich-get-richer effect. At any given $p$, RSUs maintain substantially higher variance than the network-wide average. This disparity stems from their strategic deployment at key locations (e.g., intersections), which ensures higher baseline connectivity, and the inherent heterogeneity in their spatial distribution leading to varied interaction opportunities.

\begin{figure}[t]
	\centerline{\includegraphics[width=3.2in,keepaspectratio]{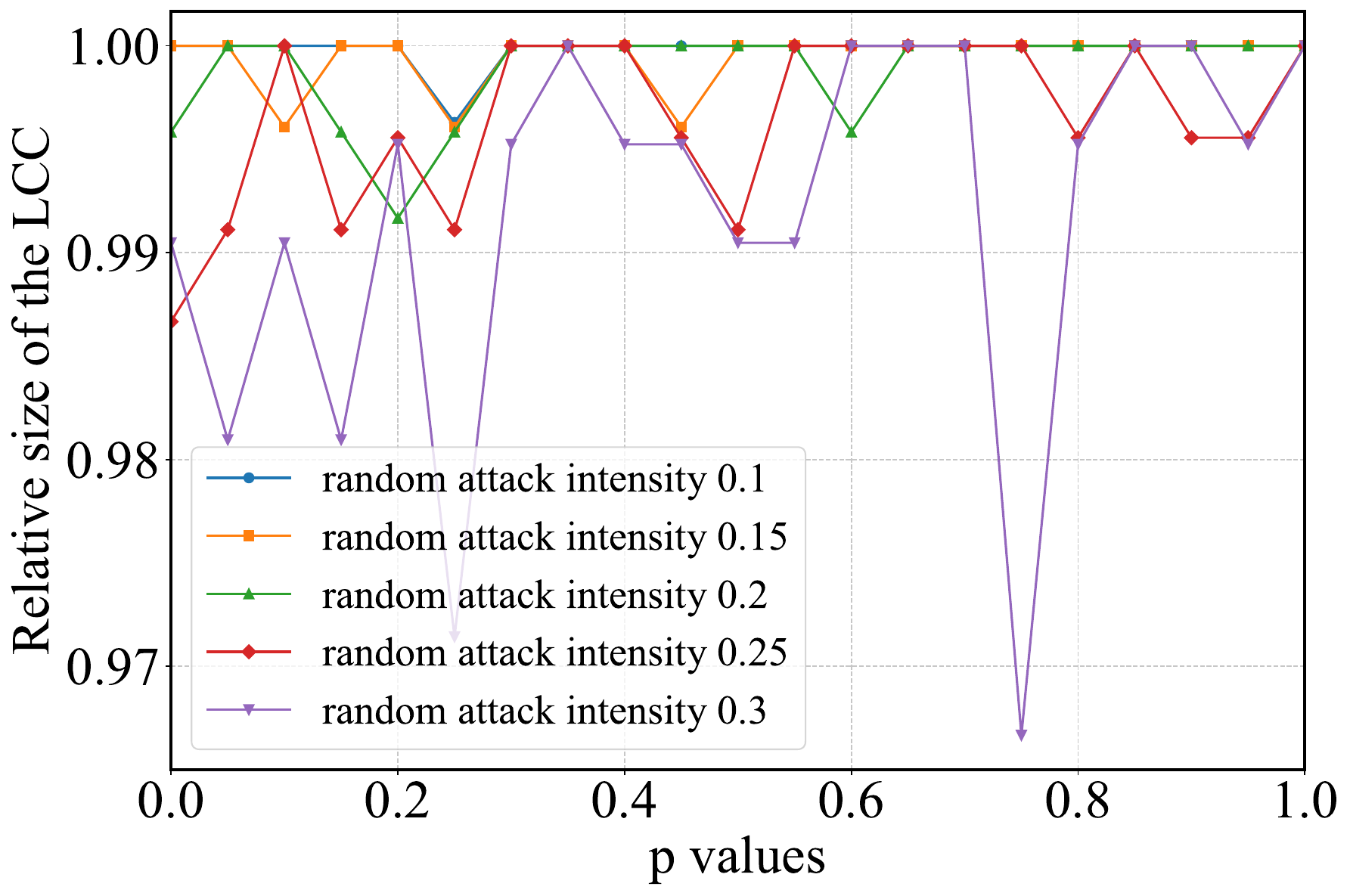}}
	\caption{Random attack trend.}
	\label{fig6}
\end{figure}

To quantify the connectivity robustness of VANETs under attack, a stable simulation step is selected, and the metric relative size of the Largest Connected Component (LCC) is introduced. This metric is defined as the ratio of the number of nodes in the largest connected component to the total number of nodes in the network. It directly quantifies the preservation of large-scale connectivity after node failures, aligning with the requirement for continuous V2V communication in VANETs and serving as a core measure of network survivability.

Fig. 6 shows the variation of the LCC size with the hybrid attachment probability $p$ under different intensities of random attacks. The results indicate that regardless of the attack intensity, the LCC size remains consistently above 0.97 and shows no significant fluctuation as $p$ changes. This stability occurs because random attacks disrupt nodes indiscriminately and are unlikely to cripple critical connective nodes. The remaining nodes can maintain large-scale connectivity through existing links, demonstrating that VANETs possess strong inherent robustness against random attacks. Their connectivity remains nearly unaffected by either the type of degree distribution or the intensity of such random failures.

\begin{figure}[t]
	\centerline{\includegraphics[width=3.2in,keepaspectratio]{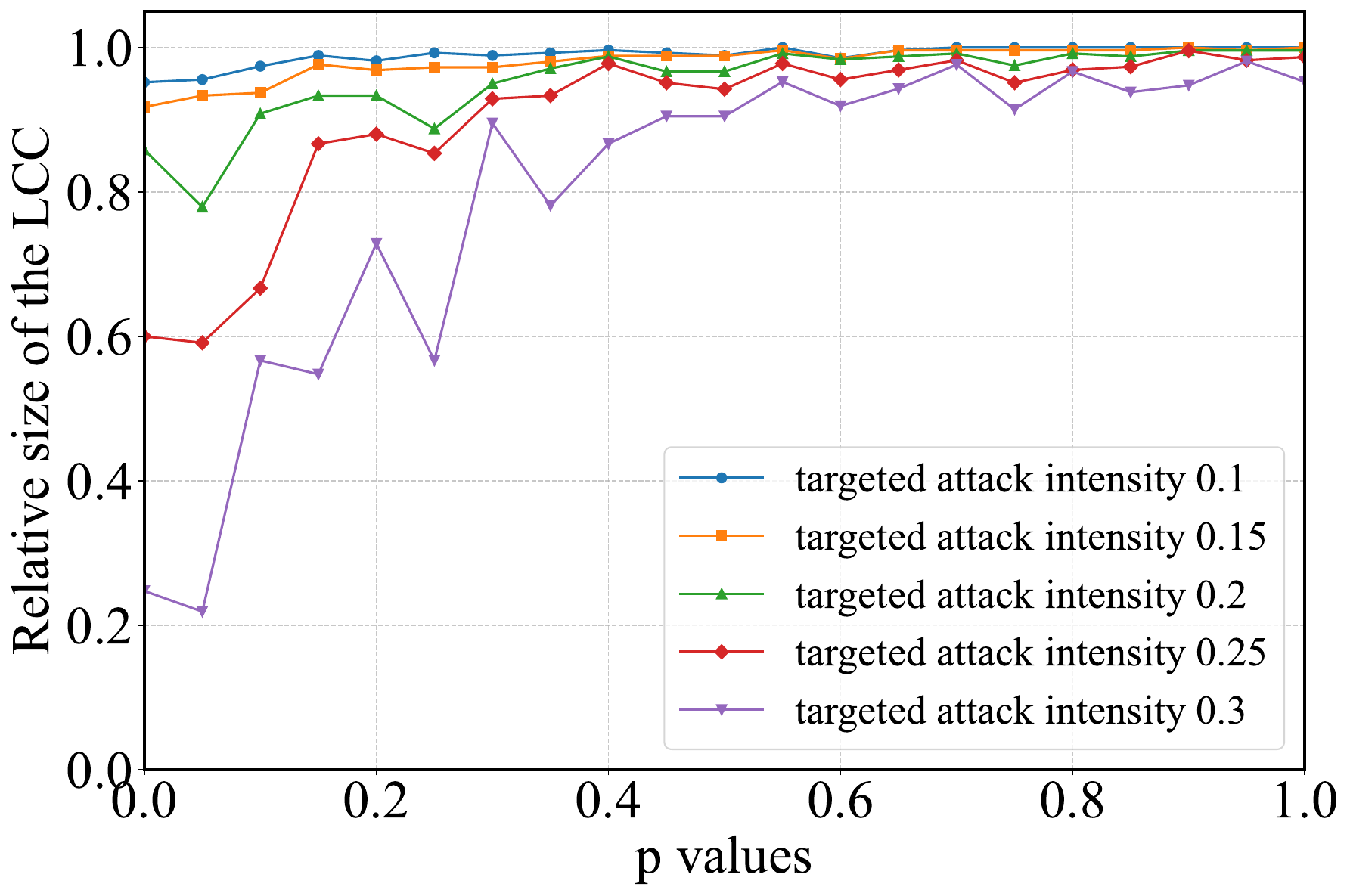}}
	\caption{Targeted attack trend.}
	\label{fig7}
\end{figure}

Fig. 7 illustrates the variation in the relative size of LCC under different intensities of targeted attacks as a function of the parameter $p$. Since targeted attacks deliberately remove high-degree nodes, their impact on connectivity strongly depends on both the attack intensity and the value of $p$. Under low-intensity attacks, only a slight fluctuation is observed near $p = 0$, without a significant drop in LCC size. This is because, even after some hub nodes are removed, the remaining hubs and numerous ordinary nodes retain sufficient alternative pathways to maintain large-scale connectivity. In contrast, under high-intensity attacks, the LCC size drops sharply at $p = 0$ but gradually recovers to nearly 1.0 as $p$ increases. This demonstrates that increasing $p$, thereby enhancing the proportion of uniform attachment, shifts the VANET from a hub-dependent architecture toward a more distributed topology, significantly improving its robustness against targeted attacks. Conversely, the power-law network at $p = 0$, which relies heavily on a few critical hubs, remains highly vulnerable to such attacks.

\begin{figure}[t]
	\centerline{\includegraphics[width=3.2in,keepaspectratio]{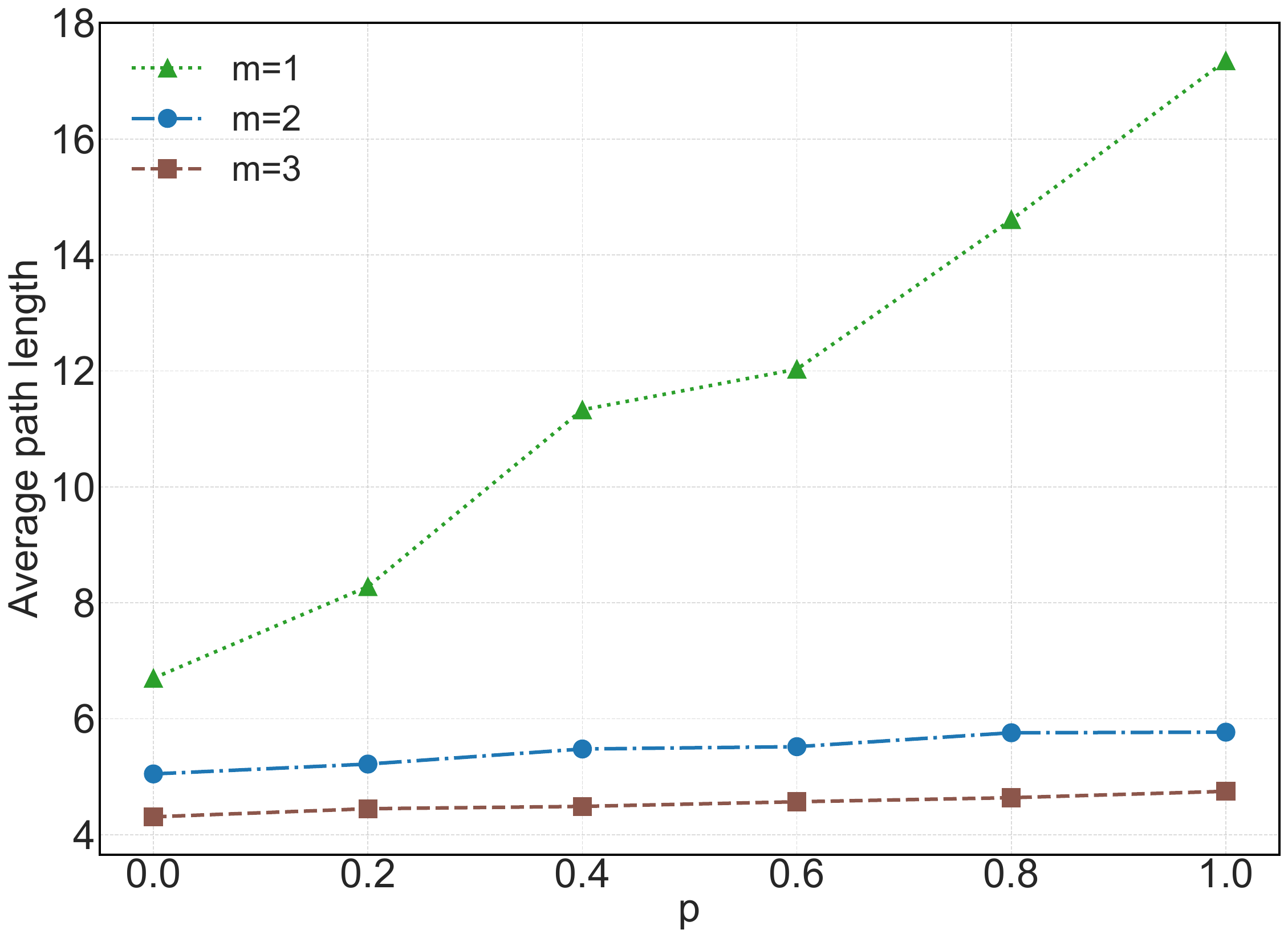}}
	\caption{Average path length of information transmission.}
	\label{fig8}
\end{figure}

Fig. 8 compares the variation of average path length with parameter $p$ under different initial connection numbers $m$ for new nodes. The results represent averages over 10 simulation steps. For all values of $m$, the average path length increases with $p$. This occurs because as $p$ increases, regulating the balance between uniform and preferential attachment, the degree distribution shifts from a pure power-law (under preferential attachment) to a tunable power-law variant, eventually approaching an exponential distribution (under uniform attachment). This progression reduces the availability of efficient hub-based routing paths, forcing information to traverse more intermediate nodes.

Notably, for $m = 2$ and $m = 3$, the average path length remains relatively short and shows low sensitivity to changes in $p$. This stability stems from higher initial connectivity: during the preferential attachment phase, multiple initial connections enhance the multi-path relay capability of hubs, maintaining short paths. Even as uniform attachment increases with $p$, the dense connectivity provides ample alternative routes, preventing significant path elongation. In contrast, the sparse network with $m = 1$ exhibits longer paths and greater sensitivity to $p$, as the limited connectivity amplifies the impact of hub dilution on routing efficiency.

\section{Conclusion and Discussion}
This paper addressed the topological governance challenges in localized VANET under dynamic traffic scenarios by proposing a schedulable degree distribution model based on a hybrid attachment mechanism. Through theoretical analysis and simulation experiments, we systematically demonstrated the model's core capability: by adjusting the hybrid attachment probability $p$, the network degree distribution can transition continuously between an exponential and a power-law distribution. This enabled dynamic trade-offs between network robustness and low-latency transmission efficiency. Simulation results confirmed that increasing $p$ effectively suppressed node degree heterogeneity and enhanced tolerance to targeted attacks, albeit at the cost of reduced transmission efficiency. Conversely, decreasing $p$ promoted hub formation, shortend the average path length to meet low-latency demands, but introduced vulnerability to hub failures. These findings validated the model's potential to provide a novel governance paradigm with a malleable underlying topology for diverse scenarios such as urban congestion and suburban highways.

However, this study has certain limitations. The core tuning parameter $p$ is currently treated as a pre-defined static value, requiring manual configuration or prior knowledge of the external environment. This open-loop control paradigm cannot adequately respond to the real-time, nonlinear variations in environmental factors like vehicle density, mobility speed, and communication load within VANETs, thus failing to fully unlock the model's potential for autonomous, closed-loop governance in realistic dynamic settings.

Looking forward, we aim to transition the model from being merely schedulable to becoming self-scheduling. A highly promising direction is the integration of Deep Reinforcement Learning (DRL) to address the aforementioned limitations. Specifically, future work will focus on developing a DRL agent that continuously perceives the network state (e.g., node density, connection dynamics, historical attack frequency) as its input, uses the hybrid attachment weighting as its action output, and operates based on a reward function designed to optimize long-term network performance (e.g., comprehensively balancing connectivity, latency, and energy consumption). This approach would enable VANETs to autonomously and intelligently adapt their topological generation strategies in real-time according to operational contexts, ultimately realizing a truly adaptive and resilient intelligent transportation network system.

\bibliographystyle{ieeetr} %为文献的格式类型
\bibliography{myref} % 为我们.bib文件名
~~~\\
~~~\\
\end{document}